\documentclass[conference,a4paper]{IEEEtran}

\usepackage[utf8]{inputenc}
\usepackage[T1]{fontenc}
\usepackage{url}              
\usepackage{cite}             

\usepackage[cmex10]{amsmath}  
\usepackage{mleftright}       
\mleftright                   

\usepackage{graphicx}         
\usepackage{booktabs}         

\usepackage{amsmath}
\usepackage{latexsym}
\usepackage{amssymb}

\newcommand{\cA}{{\cal A}}
\newcommand{\cB}{{\cal B}}
\newcommand{\cC}{{\cal C}}

\newcommand{\cF}{{\cal F}}

\newcommand{\cU}{{\cal U}}
\newcommand{\cV}{{\cal V}}

\newcommand{\bD}{{\bf D}}
\newcommand{\bE}{{\bf E}}

\newcommand{\mmod}{{\mbox{mod}}}

\newcommand{\field}[1]{\mathbb{#1}}

\newcommand{\C}{\field{C}}

\newtheorem{theorem}{Theorem}
\newtheorem{definition}{Definition}
\newtheorem{lemma}{Lemma}
\newtheorem{example}{Example}

\begin{document}

\title{Pseduo-Random and de Bruijn Array Codes}

\author{%
  \IEEEauthorblockN{Tuvi Etzion}
\IEEEauthorblockA{Dept. of Computer Science\\
Technion-Israel Institute of Technology\\
Haifa 3200003, Israel\\
email: etzion@cs.technion.ac.il}
}

\maketitle

\begin{abstract}
Pseudo-random arrays and perfect maps are the two-dimensional analogs of M-sequences and de Bruijn sequences, respectively.
We modify the definitions to be applied to codes. These codes are also the two-dimensional analogs of certain
factors in the de Bruijn graph. These factors are called zero factors and perfect factors in the de Bruijn graph.
We apply a folding technique to construct pseudo-random array codes and examine the minimum distance of the constructed codes.
The folding is applied on sequences generated from irreducible polynomials or a product of irreducible polynomials
with the same degree and the same exponent.
Direct and recursive constructions for de Bruijn array codes are presented and discussed.
\end{abstract}

\section{Introduction}
\label{sec:PM+PR}

Generalizations of one-dimensional sequences and codes to higher dimension is quite natural from
both theoretical and practical point of view. Such generalizations were considered
for various structures such as error-correcting codes~\cite{Rot91}, burst-correcting codes~\cite{EtYa09},
constrained codes~\cite{TER09}, and de Bruijn sequences~\cite{Etz88,Pat94}.
This paper considers generalizations of one-dimensional sequences with a window property to two-dimensional arrays with a window
property. For simplicity, in this paper, only binary arrays and sequences are considered, although most of the
results can be generalized to any finite field.

One-dimensional sequences are associated with {\bf \emph{the de Bruijn graph $G_n$}}, whose vertices are represented by the $2^n$~binary
words of length $n$. The graph has $2^{n+1}$ directed edges represented by the $2^{n+1}$ binary words of length $n+1$.
There is an edge from the vertex $(x_0,x_1,\ldots,x_{n-1})$ to the vertex $(x_1,\ldots,x_{n-1},x_n)$. This edge is
associated with the $(n+1)$-tuple $(x_0,x_1,\ldots,x_{n-1},x_n)$.
A Hamiltonian cycle in~$G_n$ is equivalent to a cyclic sequence of length $2^n$ in which each binary $n$-tuple is
contained exactly once as a window of length $n$. Such a sequence is equivalent to an Eulerian cycle in $G_{n-1}$.
This sequence is a {\bf \emph{span $n$ de Bruijn sequence}}. A shortened span $n$ de Bruijn sequence is a cyclic
sequence of length $2^n -1$ in which each nonzero $n$-tuple is contained exactly once in a window of length $n$.
If the shortened de Bruijn sequence can be generated by a linear recursion of degree $n$, then it is called an {\bf \textup{M}\emph{-sequence}}.
Such a sequence is generated from a recursion induced from a primitive polynomial of degree $n$.
If $S$ is an M-sequence, then $S + \bE^i S$, $1 \leq i \leq 2^n -2$, is another shift of $S$, where for any sequence~$X$,
$\bE^i X$ is a cyclic shift of a sequence $X$ by $i$ position to the left. This is the {\bf \emph{shift-and-add property}}.
The book of Golomb~\cite{Gol67} is an excellent introduction to sequences generated by irreducible polynomials and associated shift registers.
A new comprehensive book was published recently~\cite{Etz24}.
The graph $G_n$ was generalized into a two-dimensional analog~\cite{Etz88}. The two-dimensional analog of the
sequences were discussed first in~\cite{Gor66,McSl76,ReSt62}.

A {\bf \emph{factor}} in a graph is a set of vertex disjoint cycles that contain all the vertices in the graph.
A factor in $G_n$ is associated with a state diagram of a nonsingular feedback shift register.
Two types of factors are of interest to our exposition.
\begin{definition}
A {\bf \emph{perfect factor}} \textup{PF}$(n,k)$ in $G_n$ is a set of $2^{n-k}$ vertex-disjoint cycles of length $2^k$ in $G_n$.
On each cycle, one vertex ($n$-tuple) is chosen arbitrarily to be the {\bf \emph{zero state}}.
The cycles of the factor are numbered by $X_1,X_2,\ldots,X_{2^{n-k}}$.
\end{definition}

The following theorem on the existence of perfect factors was proved in~\cite{Etz88}.
\begin{theorem}
\label{thm:PF_exist}
A perfect factor \textup{PF}$(n,k)$ exists if and only if $k \leq n < 2^k$.
All sequences of a perfect factor with a given set of parameters can have the same weight parity (even or odd).
\end{theorem}

Similarly to perfect factors, we have zero factors.

\begin{definition}
A {\bf \emph{zero factor}} ZF$(n,k)$ with {\bf \emph{exponent}} $k$ in $G_n$ is a set of $d$ vertex-disjoint cycles of length $k$ in $G_n$,
which contains each nonzero $n$-tuple exactly once as a window in one of the cycles.
\end{definition}

Clearly, for a zero factor with $d$ cycles of length $k$, we must have that $d \cdot k = 2^n -1$ and $n < k \leq 2^n -1$.
All known parameters $n$, $d$, and $k$, can be inferred  from the following three theorems
proved in~\cite{Gol67} and mentioned in~\cite{Etz88}.

\begin{theorem}
\label{thm:zero_irreduc}
If the characteristic polynomial $f(x)$ of a shift register is irreducible, then the shift register
produces a zero factor with exponent $k$, where $k$ is the smallest integer such that $f(x)$ divides $x^k-1$.
\end{theorem}

\begin{theorem}
\label{thm:zero_prod_irreduc}
Let $f_i (x)$, $1 \leq i \leq r$, be $r$ different irreducible polynomials of degree $n$,
and let their corresponding shift registers have zero factors with exponent $e$.
Then the feedback shift register which has the characteristic polynomial $\prod_{i=1}^r f_i (x)$
produces a zero factor with exponent $e$.
\end{theorem}

\begin{theorem}
\label{thm:param_zero}
Every factor $k$ of $2^n-1$ which is not a factor of any number $2^\ell -1$, where $\ell < n$,
occurs as an exponent of a zero factor which corresponds to an irreducible polynomial
of degree $n$. There are $\frac{\phi (k)}{n}$ irreducible polynomials that correspond to
zero factors with exponent $k$, where $\phi$ is the Euler totient function.
\end{theorem}

It is an open intriguing problem to find new zero factors with new parameters or zero factors with the same set
of parameters, but not obtained from Theorems~\ref{thm:zero_irreduc},~\ref{thm:zero_prod_irreduc}, or~\ref{thm:param_zero}.
The goal of this paper is to discuss two-dimensional generalizations of zero factors and perfect factors.
These two-dimensional generalizations will be called pseudo-random array codes and de Bruijn array codes, respectively.
They are motivated from both practical an theoretical point of views.

The rest of this paper is organized as follows. In Section~\ref{sec:preliminaries} we present the basic definitions and results for
the arrays that will be considered in this work.
A construction for two-dimensional pseudo-random array codes will be given in Section~\ref{sec:PR_array_codes}.
The construction is based on the folding of one-dimensional sequences into appropriate arrays.
Sufficient and necessary conditions that the folding will form the required arrays will be discussed.
Two-dimensional perfect-maps codes will be discussed in Section~\ref{sec:2Dperfect_factor}. These arrays
can be viewed as two-dimensional perfect factors. Direct constructions for such arrays will be presented in this section
as well as a few recursive constructions. Section~\ref{sec:conclusion} includes a conclusion and future work. For lack of space many
results were given with no proof.

\section{Preliminaries}
\label{sec:preliminaries}

In this section, we provide the basic definitions and results of two-dimensional
arrays and two-dimensional array codes that are considered.
All the arrays that will be considered in this paper will be $r \times t$ cyclic (doubly-periodic) arrays, i.e.,
the array is read cyclically, vertically and horizontally.

\begin{definition}
A {\bf \emph{perfect map}} (or a {\bf \emph{de Bruijn array}}) is an $r \times t$ doubly-periodic array,
such that each $n \times m$ matrix appears exactly once
as a window in the array. Such an array will be called an $(r,t;n,m)$-PM or an $(r,t;n,m)$-DBA.
\end{definition}

\begin{definition}
A {\bf \emph{shortened perfect map}} (or a {\bf \emph{shortened de Bruijn array}}) is an $r \times t$
doubly-periodic array, such that each nonzero $n \times m$ matrix appears exactly
once as a window in the array. Such an array will be called an $(r,t;n,m)$-SPM or and $(r,t;n,m)$-SDBA.
\end{definition}

Perfect maps were first presented in~\cite{ReSt62} and later in~\cite{Gor66}. The first constructions for perfect maps
were discussed in~\cite{FFMS85,Ma84} and later in~\cite{Etz88} who also presented a construction for shortened perfect maps.

\begin{lemma}
\label{lem:condPM}
If $A$ is an $(r,t;n,m)$-$\textup{PM}$ then
\begin{enumerate}
\item[{\bf 1.}] $r > n$ or $r=n=1$,

\item[{\bf 2.}] $t > m$ or $t=m=1$, and

\item[{\bf 3.}] $rt = 2^{nm}$.
\end{enumerate}
\end{lemma}

\begin{lemma}
\label{lem:condSPM}
If $A$ is an $(r,t;n,m)$-$\textup{SPM}$ then
\begin{enumerate}
\item[{\bf 1.}] $r > n$ or $r=n=1$,

\item[{\bf 2.}] $t > m$ or $t=m=1$, and

\item[{\bf 3.}] $rt = 2^{nm} -1$.
\end{enumerate}
\end{lemma}
Paterson~\cite{Pat94} proved that the necessary conditions of Lemma~\ref{lem:condPM} are also sufficient.
\begin{definition}
A {\bf \emph{pseudo-random array}} $\cA$ is an $(r,t;n,m)$-SPM
such that $\cA + \cA'$, where $\cA'$ is a nontrivial shift of $\cA$, is another nontrivial shift of $\cA$.
Such an array will be called an $(r,t;n,m)$-PRA and it has the shift-and-add property.
\end{definition}

Pseudo-random arrays were constructed first in~\cite{Spa63} and later in~\cite{McSl76,NMIF72}.
In the current work, we are interested
in constructing a set of arrays with the same size for which each $n \times m$ matrix, with the possible exception
of the all-zeros matrix, is contained in exactly one of the arrays from the set.

\begin{definition}
A {\bf \emph{de Bruijn array code}} is a set of size $\ell$ of $r \times t$ doubly-periodic arrays,
such that each $n \times m$ matrix appears exactly once
as a window in one of the arrays. Such a set of arrays will be called an $(r,t;n,m)$-DBAC.
\end{definition}
Note, that a de Bruijn array code is the two-dimensional analog of a perfect factor.
\begin{lemma}
\label{lem:condPMC}
If $\C$ is an $(r,t;n,m)$-$\textup{DBAC}$ of size $\ell$, then
\begin{enumerate}
\item[{\bf 1.}] $r > n$ or $r=n=1$,

\item[{\bf 2.}] $t > m$ or $t=m=1$.

\item[{\bf 3.}] $\ell rt = 2^{nm}$.
\end{enumerate}
\end{lemma}

\begin{definition}
A {\bf \emph{shortened de Bruijn array code}} is a set of size $\ell$ of $r \times t$ doubly-periodic arrays,
such that each nonzero $n \times m$ matrix appears exactly once
as a window in one of the arrays.
Such a set of arrays will be called an $(r,t;n,m)$-SDBAC.
\end{definition}
Note, that a shortened de Bruijn array code is the two-dimensional analog of a zero factor.
\begin{definition}
A {\bf \emph{pseudo-random array code}} is a shortened de Bruijn array code $\C$ of size $\ell$ with $r \times t$ doubly-periodic arrays,
with the following property.
Given $\cA \in \C$, and $\cA' \in \C$, where either $\cA$ and $\cA'$ are distinct arrays or $\cA'$ is a nontrivial shift of $\cA$,
then $\cB=\cA + \cA'$, where $\cB \in \C$.
Such a set of arrays will be called an $(r,t;n,m)$-PRAC and together they have the shift-and-add property.
These arrays are the analogs of the factors constructed in Theorems~\ref{thm:zero_irreduc},~\ref{thm:zero_prod_irreduc},
and~\ref{thm:param_zero}.
\end{definition}

\section{Pseudo-Random Array Codes}
\label{sec:PR_array_codes}
We start with a construction presented in MacWilliams and Sloane~\cite{McSl76}.
Assume that $\eta=2^{k_1k_2}-1$, $r =2^{k_1}-1$, and $t = \frac{\eta}{r}$,
where ${\text{g.c.d.}(r,t)=1}$.
Let $S=s_0s_1s_2 \cdots$ be a span $k_1 k_2$ M-sequence. Write $S$ down the right diagonals of an $r \times t$ array $B=\{ b_{ij}\}$,
$0 \leq i \leq r -1$, $0 \leq j \leq t -1$, starting
at $b_{00} , b_{11} , b_{22}$ and so on, where the last position is $b_{r -1,t -1}$.
After $b_{ij}$ we continue to write $b_{i+1,j+1}$, where $i+1$ is taken modulo $r$ and $j+1$
is taken modulo~$t$. This method is called {\bf \emph{folding}}. The following theorem was proved in~\cite{McSl76}.

\begin{theorem}
\label{thm:PRA_MS}
Each $k_1 \times k_2$ nonzero matrix is contained exactly once as a window in the $r \times t$ array~$B$, where $r = 2^{k_1}-1$
and $t =(2^{k_1k_2}-1)/{r}$, i.e., $B$ is an $(r,t;k_1,k_2)$-PRA.
\end{theorem}

\begin{example}
\label{ex:foldPR}
For $k_1=k_2=2$, $r=3$, and $t=5$,
consider the span $4$ ${\textup{M}\text{-sequence}}$ $S=[000111101011001]$, with positions numbered
from $0,1$, up to~$14$. Consider now the $3 \times 5$ array $B$ with the entries~$b_{ij}$,
$0 \leq i \leq 2$, $0 \leq j \leq 4$, where the positions $0$ through $14$, of the sequence, are folded into $B$ as follows
$$
\left[
\begin{array}{ccccc}
b_{00} & b_{01} & b_{02} & b_{03} & b_{04} \\
b_{10} & b_{11} & b_{12} & b_{13} & b_{14} \\
b_{20} & b_{21} & b_{22} & b_{23} & b_{24}
\end{array}
\right],~
\left[
\begin{array}{ccccc}
0 & 6 & 12 & 3 & 9 \\
10 & 1 & 7 & 13 & 4 \\
5 & 11 & 2 & 8 & 14
\end{array}
\right] .
$$
The $\textup{M}$-sequence $S$ is folded into the array $B$ keeping the order of the entries in~$S$ according to the order defined by $B$, i.e.,
$$
\left[
\begin{array}{ccccc}
0 & 1 & 0 & 1 & 0 \\
1 & 0 & 0 & 0 & 1 \\
1 & 1 & 0 & 1 & 1
\end{array}
\right]
$$
which forms a $(3,5;2,2)$-$\textup{PRA}$.
\hfill\quad $\blacksquare $
\end{example}

The important requirement used in Theorem~\ref{thm:PRA_MS} to prove that the array $B$ is an $(r,t;k_1,k_2)$-PRA is that
$r=2^{k_1}-1$. Is this requirement necessary? It appears that this requirement is necessary in some cases, but it is not required
in other cases. The proof of Theorem~\ref{thm:PRA_MS} given in~\cite{McSl76} is based on the observation that in the
top $k_1 \times k_2$ window of the array $A$ we cannot have the all-zeros $k_1 \times k_2$ matrix. For this purpose
the simplex code of length $2^{k_1k_2}-1$ and dimension $k_1 k_2$ is used. Now, we will provide an alternative proof that will apply
also to pseudo-random array codes. For this purpose, we will develop some simple theory. The concepts of this theory
are similar to the ones developed for VLSI testing in Lempel and Cohn~\cite{LeCo85}, but the main proof is different as the
proof in~\cite{LeCo85} does not hold for the generalized theory which will be presented.

\begin{definition}
For a set $R= \{r_0,r_1,\ldots,r_{t-1} \}$ of $t$ positions in a sequence, the {\bf \emph{set polynomial}} $g_R(x)$
is defined by
$$
g_R(x) \triangleq \prod_{Q \subseteq R} \sum_{r_i \in Q} x^{r_i}
$$
\end{definition}

Let $f(x) = 1 +\sum_{i=1}^n c_i x^i$ be characteristic irreducible polynomial and its associated sequence (or several sequences)
$A=a_0a_1a_2a_3 \cdots$ (an M-sequence if the polynomial is primitive and a few nonzero sequences, of the same length, if the
polynomial is irreducible and not primitive) which satisfies the recurrence
\begin{equation}
\label{eq:recVLSI}
a_k = \sum_{i=1}^n c_i a_{k-i}
\end{equation}
with the initial nonzero $n$-tuple $(a_{-n} a_{-n+1} ~ \cdots ~ a_{-1})$.

Consider all the possible shifts of the nonzero sequences generated by $f(x)$ as rows in a matrix $B$, which has $L=2^n-1$ rows, and
let $T$ be the $L \times n$ matrix which is formed by a projection of any $n$ columns of $B$.
Note, that by Theorem~\ref{thm:zero_irreduc} all the nonzero sequences have the same period and hence there is no ambiguity.

\begin{lemma}
\label{lem:tupleA}
Every nonzero $n$-tuple appears as a row of the matrix $T$ if and only if the columns of $T$ are linearly independent.
\end{lemma}
\begin{IEEEproof}
Assume first that each $n$-tuple appears as a row in $T$. This immediately implies that the $n$ columns of $T$ are linearly independent.

Assume now that the columns of $T$ are linearly independent.
Since each $n$-tuple appears as a window exactly once in one of the nonzero sequences generated by $f(x)$, it follows
that every $n$ consecutive columns of $B$ contain each one of the $2^n-1$ nonzero $n$-tuples as a row.
Hence, the first $n$ columns of $B$ contain each nonzero $n$-tuple exactly once.
These column vectors can be used as rows for the generator matrix of the simplex code of length $2^n-1$ and dimension $n$.
Each other column of $B$ can be represented as a linear combination of the first $n$ columns of $B$.
This linear combination is defined by the recursion induced by $f(x)$ given in Eq.~(\ref{eq:recVLSI}).
Hence, all these linear combinations coincide with the codewords of the simplex code.
This implies that every $n$ linearly independent columns contain each nonzero $n$-tuple as a row in $T$.
\end{IEEEproof}

\begin{lemma}
\label{lem:pqA}
If $Q$ is a nonempty subset of $R$ and $q(x) = \sum_{r_i \in Q} x^{r_i}$, then $f(x)$~divides $q(x)$
if and only if the columns of $B$ that are associated with the subset~$Q$ sum to \emph{zero}.
\end{lemma}
\begin{IEEEproof}
If the columns in $B$ that are associated with the subset $Q$ sum to \emph{zero}, then one of the columns
is a sum of the other columns, i.e., this column is a linear combination of the other columns.
This linear combination is induced by the polynomial $f(x)$ and hence $q(\beta)=0$, where $\beta$
is a root of~$f(x)$. Since we also have $f(\beta)=0$ and $f(x)$ is an irreducible polynomial, it follows that $f(x)$~divides $q(x)$.

If $f(x)$ divides $q(x)$, then $f(\alpha)=0$ implies that $q(\alpha)=0$ and hence since the columns of $B$ are defined
by the recursion induced by $f(x)$, it follows that associated columns of $B$ defined by $Q$ sum to \emph{zero}.
\end{IEEEproof}

\begin{theorem}
\label{thm:px_gRx}
Given an irreducible polynomial $f(x)$ and a set~$R$ of $n$ coordinates in $B$, then the set of~$R$ coordinates in~$B$ contains each nonzero
$n$-tuple if and only if $g_R (x)$ is not divisible by~$f(x)$.
\end{theorem}
\begin{IEEEproof}
Consider the $L \times n$ matrix $T$ projected by the $n$ columns of $B$ associated with the coordinates of $R$.
By Lemma~\ref{lem:tupleA} every nonzero $n$-tuple appears as a row in $T$ if and only if the columns of $T$
are linearly independent. The columns of $T$ are linearly dependent if and only if
a nonempty subset of the columns in $T$ sums to \emph{zero}.

By Lemma~\ref{lem:pqA} we have that $f(x)$ divides the polynomial $\sum_{r_i \in Q} x^{r_i}$,
where $Q$ is a nonempty subset of $R$, if and only if the associated subset of columns
of~$T$ sums to \emph{zero}.

Since the polynomial $f(x)$ is irreducible, it follows that $f(x)$ divides the set polynomial $g_R(x)$ if and only if
there exists a subset $Q \subseteq R$ such that $f(x)$~divides the factor $\sum_{r_i \in Q} x^{r_i}$
of $g_R(x)$. Hence, by Lemmas~\ref{lem:tupleA} and~\ref{lem:pqA} the proof is completed.
\end{IEEEproof}

Theorem~\ref{thm:px_gRx} yields a method to verify whether a set of arrays formed by folding forms
an $(r,t;m,n)$-PRAC.

A horizontal shift and/or a vertical shift of the array will be equivalent to an array obtained by folding
from another point of the M-sequence.
Since the ${\text{M-sequence}}$ has the shift-and-add property, it follows that if we make
any such shift, the two arrays will sum to another
shift of the array. This is the shift-and-add property of the array.

\begin{example}
\label{ex:fold_PM}
Consider the array and the $\textup{M}$-sequence~$S$ of Example~\ref{ex:foldPR}. We shift the array horizontally by $2$ and vertically
by $1$ and add them as follows, where the first bit of $S$ is in bold.
$$
\left[
\begin{array}{ccccc}
{\bf 0} & 1 & 0 & 1 & 0 \\
1 & 0 & 0 & 0 & 1 \\
1 & 1 & 0 & 1 & 1
\end{array}
\right] +
\left[
\begin{array}{ccccc}
1 & 1 & 1 & 1 & 0 \\
1 & 0 & {\bf 0} & 1 & 0 \\
0 & 1 & 1 & 0 & 0
\end{array}
\right]
$$
$$
=\left[
\begin{array}{ccccc}
1 & 0 & 1 & 0 & {\bf 0} \\
0 & 0 & 0 & 1 & 1 \\
1 & 0 & 1 & 1 & 1
\end{array}
\right] .
$$
The \textup{M}-sequence $S$ starts in the leftmost array in $b_{00}$, in the middle array at $b_{12}$, and in the rightmost array at $b_{04}$.
\hfill\quad $\blacksquare $
\end{example}

The folding construction of MacWilliams and Sloane~\cite{McSl76}
can be generalized to obtain array codes with large minimum distance, such that each nonzero array
of a certain size is contained in exactly one of the codewords as a window. Instead of folding as M-sequence
as done at the beginning of this section, we consider all the nonzero sequences generated by an irreducible polynomial
and use either Theorem~\ref{thm:px_gRx} or some theorems which follow.
There are a few constructions of pseudo-random array codes which are based either
on folding the sequences generated by an irreducible polynomial or by folding the sequences generated by a product
of a few irreducible polynomial of the same degree and the same exponent.

\begin{theorem}
\label{thm:PRA_Codes}
Let $f(x)$ be an irreducible polynomial of degree~$nm$ with exponent $e$, i.e., $2^{nm} -1 = k \cdot e$.
Assume further that $e$ can be factorized to $e = (2^n -1) \ell$.
If g.c.d.$(2^n-1,\ell)=1$, then we can fold the $k$ cycles generated by $f(x)$ into $(2^n-1) \times \ell$ arrays.
Let $R$ be the set of positions of a cycle folded into an array located in any $n \times m$ sub-array.
If $f(x)$ does not divide $g_R (x) = \prod_{Q \subseteq R} \sum_{r_i \in Q} x^{r_i}$,
then the $k$ arrays form a $(2^n-1,\ell;n,m)$-PRAC.
\end{theorem}

There are some important cases when Theorem~\ref{thm:px_gRx} is not required and they are given in the following theorems.
\begin{theorem}
\label{thm:newA1}
Let $f(x)$ be an irreducible polynomial of degree~$mn$ with exponent $(2^n-1)(2^m-1)$.
Then the cycles generated by $f(x)$ are folded into $(2^n -1) \times (2^m -1)$ arrays and
yield a $(2^n-1,2^m-1;n,m)$-PRAC.
\end{theorem}

\begin{theorem}
\label{thm:newB2}
Let $f(x)$ be an irreducible polynomial of degree~$\rho$ and exponent $r$. Let $g(x)$ be an irreducible polynomial of degree $\tau$,
and exponent $t$, where ${\text{g.c.d.}(r,t)=1}$. Consider now a sequence $\cV = v_0 v_1 \ldots v_{r-1}$
generated by $f(x)$ and a sequence $\cU=u_0u_1 \ldots u_{t-1}$
generated by $g(x)$. Let $\cC$ be an $r \times t$ doubly-periodic array whose $i,j$ entry, $0 \leq i \leq r-1$, $0 \leq j \leq t-1$
is $v_i \cdot u_j$. Then the linear span of the array $\cC$ forms an $(r,t;\rho , \tau)$-PRAC.
\end{theorem}

\begin{example}
\label{ex:3x7arrays}
Consider the characteristic irreducible polynomial
$$
f(x)=  x^6 +x^5 + x^4 + x^2 +1
$$
and its associated feedback (recursive) function
$$
x_7 = f(x_1,x_2,x_3,x_4,x_5,x_6) = x_1 + x_2 + x_3 + x_5 ~.
$$
The polynomial $f(x)$ generates three nonzero sequences
$$
[{\bf 0}00001010010011001011],
$$
$$
[{\bf 0}10101110100001111011],
$$
$$
[{\bf 1}11100111000100011011].
$$
Folding these three sequences into a $3 \times 7$ arrays yield the following three arrays
$$
\left[
\begin{array}{ccccccc}
{\bf 0} & 0 & 0 & 0 & 0 & 0 & 0 \\
1 & 0 & 0 & 1 & 0 & 1 & 1 \\
1 & 0 & 0 & 1 & 0 & 1 & 1
\end{array}
\right] , ~
\left[
\begin{array}{ccccccc}
{\bf 0} & 1 & 1 & 1 & 0 & 0 & 1 \\
1 & 1 & 1 & 0 & 0 & 1 & 0 \\
1 & 0 & 0 & 1 & 0 & 1 & 1
\end{array}
\right] ,
$$
and
$$
\left[
\begin{array}{ccccccc}
{\bf 1} & 0 & 0 & 1 & 0 & 1 & 1 \\
1 & 1 & 1 & 0 & 0 & 1 & 0 \\
0 & 1 & 1 & 1 & 0 & 0 & 1
\end{array}
\right]  ~.
$$
These three arrays form a $(3,7;2,3)$-PRAC with minimum distance~8.
The related code of length 21 and dimension 6 has at most minimum distance 8, i.e., the constructed code is optimal.
\hfill\quad $\blacksquare $
\end{example}
Generally, the minimum distance of the PRAC is the weight of the sequence of the smallest
weight generated by $f(x)$.

\section{Two-Dimensional de Bruijn Array Codes}
\label{sec:2Dperfect_factor}

In this section, we present several direct and recursive constructions for de Bruijn array codes.
These are also two-dimensional perfect factors since they form a factor with cycles of the same length in a
two-dimensional generalization of the de Bruijn graph~\cite{Etz88}. The first type of arrays are obtained by concatenating
sequences from a perfect factor.

\begin{theorem}
An $(2^k,2^s; n,1)$-DBAC, $s \geq 1$, of size $\ell$ exists if and only if there exists a PF$(n,k)$, where $2^{n-k}=\ell \cdot 2^s$.
\end{theorem}

\begin{theorem}
If there exist a PF$(n,n-\ell)$, where $1 \leq \ell$, then there exists a $(2^{n-\ell},2^m;n,2)$-DBAC, where $2 \leq m \leq n-\ell +1$.
\end{theorem}


\begin{theorem}
\label{thm:PMC_odd}
Let $\cF$ be a PF$(n,k)$, $\ell = 2^m-1$, $m \geq k$, and define the following set of matrices
$$
\C \triangleq \{ [X_{i_1},\bE^{j_2} X_{i_2},\ldots,\bE^{j_\ell}X_{i_\ell},\bE^{j_{\ell+1}} X_{i_{\ell+1}}] ~:~
$$
$$
X_{i_r} \in \cF, ~\sum_{r=1}^{\ell+1} i_r \equiv 1 ~(\mmod ~ 2^{n-k}),~ \sum_{r=2}^{\ell+1} j_r \equiv 0 ~(\mmod ~ 2^k) \},
$$
where the $X_{i_r}$'s are column vectors and their cyclic shifts are related to their zero states.
Then, the code $\C$ is a $(2^k,2^m;n,2^m -1)$-DBAC of size $2^{n \ell -k-m}$.
\end{theorem}
\begin{IEEEproof}
The codewords of the construction are $2^k \times 2^m$ matrices, but they can also be viewed as cycles of length $2^m$.
We start by showing that each cycle defined in $\C$ has length ${\ell +1 = 2^m}$.
If the length of some cycle $\cC$ in $\C$ is smaller than $2^m$, then
it should be a divisor of $2^m$, i.e., a power of 2. Assume the contrary that this period is $2^s$ for some $s$, ${0 \leq s < m}$.
Since $\sum_{r=1}^{\ell+1} i_r \equiv 1 ~(\mmod ~ 2^{n-k})$, it follows that $\sum_{r=1}^{\ell+1} i_r = \alpha 2^{n-k} +1$ and hence
$\sum_{r=1}^{2^s} i_r = \frac{\alpha 2^{n-k} +1}{2^{m-s}}$. But, $\frac{\alpha 2^{n-k} +1}{2^{m-s}}$
is not an integer, a contradiction. Therefore, all the cycles of $\C$ form $2^k \times 2^m$ arrays.

We continue by computing the size of the code $\C$. Each $X_{i_r}$, $1 \leq r \leq \ell$, is taken arbitrarily from $\cF$
and hence it can be chosen in $2^{n-k}$ distinct ways. $X_{i_{\ell+1}}$ is determined
by the equation $\sum_{r=1}^{\ell+1} i_r \equiv 1 ~(\mmod ~ 2^{n-k})$. The first cycle $X_{i_1}$ is taken in its zero state,
while the next $\ell -1$ cycles can be taken in any of their $2^k$ cyclic shifts. The shift of the last cycle~$X_{i_{\ell+1}}$ is determined
by the equation $\sum_{r=2}^{\ell+1} j_r \equiv 0 ~(\mmod ~ 2^k) \}$.
Hence, each one of the middle $\ell -1$ cycles has $2^{n-k} 2^k=2^n$ possibilities.
The first cycle has $2^{n-k}$ choices and the last cycle and its shift are determined by the first $\ell$ cycles and their shifts.
Therefore, there are a total of $2^{n(\ell-1)}2^{n-k}=2^{n\ell -k}$ distinct choices for all the cycles and their shifts.

Now, we calculate the number of cycles which are counted more than once in this enumeration. Each constructed cycle~$\cC$ can start with
any one of its $\ell +1 = 2^m$ columns. This column will be taken in the shift where its zero state is the first $n$-tuple.
The other columns are defined by the order of the columns in $\cC$ and their shifts are
taken exactly as in $\cC$ related to the first state of the new first column. Therefore, each cycle $\cC$ is constructed exactly $2^m$
times in the construction of $\C$. Thus, the size of $\C$ is $2^{n\ell -k}/2^m =2^{n \ell -k-m}$.

Therefore, the number of $n \times (2^m-1)$ windows in the all the codewords of
$\C$ is $2^{n \ell -k-m} \cdot 2^k \cdot 2^m=2^{n \ell} =2^{n (2^m-1)}$.
Hence, to complete the proof it is sufficient to show that each $n \times (2^m-1)$ matrix appears as a window in a codeword of~$\C$.
Consider such a matrix $A$. The columns of $A$ are $n$-tuples in cycles of $\cF$ and
hence their associated cycles of $\cF$ can be arranged in a codeword~$\cC$
to form the window with $A$ since there are no constraint in the construction for the first $2^m-1$ columns.

Thus,  $\C$ is a $(2^k,2^m;n,2^m -1)$-DBAC of size $2^{n \ell -k-m}$.
\end{IEEEproof}

Note that in the construction of Theorem~\ref{thm:PMC_odd},
the importance of $\ell +1$ being $2^m$ and $\sum_{r=1}^{\ell+1} i_r \equiv 1 ~(\mmod ~ 2^{n-k})$ is
to avoid any horizontal periodicity in the array. The importance of $m \geq k$ is that also after a cyclic
shift vertically of the array the sum of the cyclic shifts of the columns will be zero. This is
guaranteed since the shifts are taken modulo $2^k$ and $2^k$ divides~$2^m$.

\begin{theorem}
\label{thm:PMC_SD}
Let $\cF$ be a PF$(n,k)$, $\ell = 2^m$, where $\cF$ does not contain self-dual sequences,
for a cycle $\cC \in \cF$ also $\bar{\cC} \in \cF$, and define the following code
$$
\C \triangleq \{ [X_{i_1},\bE^{j_2} X_{i_2},\ldots,\bE^{j_\ell} X_{i_\ell},
$$
$$
\bar{X}_{i_1},\bE^{j_2} \bar{X}_{i_2},\ldots,\bE^{j_\ell}\bar{X}_{i_\ell} ] ~:~
$$
$$
X_{i_r} \in \cF, ~ 0 \leq j_r \leq 2^k-1 \},
$$
where the $X_r$'s are column vectors and their cyclic shifts are related to their zero states.
Then, the code $\C$ is a $(2^k,2^{m+1};n,2^m)$-DBAC of size $2^{n \ell -k-m-1}$.
\end{theorem}

Note that in the construction of Theorem~\ref{thm:PMC_SD},
the importance of $\ell$ being $2^m$ as other value of $\ell$ implies periodicity of the array.
When $\ell = 2^m$ the codeword $\cC \in \C$ contains a sequence of cycles from $\cF$ that behave like a self-dual sequence
which cannot be horizontally periodic if its length is a power of~2.

The next set of DBACs are based on the existence of perfect factor over an alphabet of size $2^n$~\cite{Pat95}.

\begin{theorem}
If $m+1 < 2^k \leq 2^{nm}$, then there exists a $(2^n,2^k;n,m+1)$-DBAC.
\end{theorem}

For a recursive constructions, we are given an $(r,t;n,m)$-DBAC $\C$ with some parameters that will be specified
in the constructions and a perfect factor with certain parameters which depend on the parameters of $\C$.
The sequences of the perfect factor are used as indicators associated with the columns of the arrays of $\C$,
from which we decide what to do with each column in the recursion. The constructions which
are used are very similar to the ones used in~\cite{FFMS85}.

\begin{theorem}
\label{thm:DB_PMC_direct}
If there exists an $(r,2^m;n,m)$-DBAC $\C$ of size~$k$, whose columns are sequences of period~$r$ and even weight, then there
exists an $(r,2^m;n+1,m)$-DBAC $\C'$ of size $2^m k$.
\end{theorem}

The idea behind the proof of Theorem~\ref{thm:DB_PMC_direct} is to use a de Bruijn sequence $S$ of length $2^m$ and an operator $\bD$
(see~\cite{Etz88,FFMS85,Pat94}).
known as the derivative or the $\bD$-morphism. The de Bruijn sequence indicates how the inverse of $\bD$ is applied on the column
vectors of the codewords of $\C$. There are two ways to apply this inverse and the bits of $S$ indicate which one will be applied.
The sequence $S$ is taken for this purpose in all its cyclic shifts.

\begin{theorem}
\label{thm:PF_DBAC_direct}
If there exists an $(r,2^{m-\ell};n,m)$-DBAC of size $k$, where $m \leq 2^{m-\ell}$, whose columns are sequences
of period $r$ and even weight, then there exists an $(r,2^{m-\ell};n+1,m)$-DBAC of size $2^m k$.
\end{theorem}

\begin{theorem}
\label{thm:DB_even}
If there exists an $(r,t;n,m)$-DBAC of size $k$ whose columns are sequences of period $r$ and even weight, then there
exists an $(r,2^m t;n+1,m)$-DBAC of size $k$.
\end{theorem}

\begin{theorem}
\label{thm:DB_even_half}
If there exists an $(r,t;n,m)$-DBAC of size $k$ whose columns are sequences of period $r$ and even weight, then there
exists an $(r,2^{m-1} t;n+1,m)$-DBAC of size $2k$.
\end{theorem}

The next two theorems are analogs of Theorems~\ref{thm:DB_PMC_direct},~\ref{thm:PF_DBAC_direct},~\ref{thm:DB_even},
and~\ref{thm:DB_even_half} to the case where the columns of codewords in the de Bruijn array code have odd weights.


\begin{theorem}
\label{thm:PFodd_DBAC_direct}
If there exists an $(r,2^{m-\ell};n,m)$-DBAC of size $k$, where $m \leq 2^{m-\ell}$, whose columns are sequences of odd weight,
then there exists an $(2r,t;n+1,m)$-DBAC of size $2^{m-1} k$.
\end{theorem}

\begin{theorem}
If there exists an $(r,t;n,m)$-DBAC of size $k$ whose columns are sequences of odd weight, then there
exists an $(2r,2^{m-1}t;n+1,m)$-DBAC of size $k$.
\end{theorem}

\section{Conclusion and Future Work}
\label{sec:conclusion}

We have defined the concepts of pseudo-random array codes and de Bruijn array codes.
Constructions for pseudo-random array codes based on the folding of sequences generated by an irreducible polynomial,
which is not primitive or a product of irreducible polynomials of the same degree and the same exponent, is presented.
Direct constructions for de Bruijn array codes and several recursive constructions are presented.
Many questions remained unsolved and are offered for future research.

\begin{enumerate}
\item Provide constructions (with new parameters) for shortened de Bruijn arrays and
in particular pseudo-random arrays with $n \times m$ window property, where $n \geq 3$.

\item Are the necessary conditions of Lemma~\ref{lem:condSPM} sufficient?

\item Provide new constructions for zero factors which are not constructed by Theorem~\ref{thm:zero_irreduc} and~\ref{thm:zero_prod_irreduc}
and especially with parameters that are not covered by Theorem~\ref{thm:param_zero}.

\item Are the necessary conditions for the existence of a de Bruijn array codes (see Lemma~\ref{lem:condPMC}) also sufficient?

\item When folding yields a pseudo-random array, except for the set of parameters given in Theorem~\ref{thm:PRA_MS}?

\item When folding yields a pseudo-random array code, except for the set of parameters given in Theorems~\ref{thm:PRA_Codes}.
\ref{thm:newA1}, and~\ref{thm:newB2}?

\item When the condition of Theorem~\ref{thm:PRA_Codes} are satisfied?
\end{enumerate}

More properties, constructions of codes, their analysis with the recursive constructions,
and all the unproved results in this draft will be presented in~\cite{Etz24a} and~\cite{Etz24b}.

\section*{Acknowledgment}
This research was supported in part by Israel Science Foundation grant no. 222/19.
The author would like to thank Ronny Roth and Huimin Lao for their help.



\end{document}